\documentclass[preprint,proceedings]{rmaa}

\suppressfulladdresses 

\usepackage{paralist}
\usepackage{xspace}
\usepackage{psfrag,color}

\def\kms{km s$^{-1}$\xspace}

\def\tx{T$_X$\xspace}

\SetYear{2005}
\SetConfTitle{Latin American Regional IAU Meeting}

\title{The dynamical state of massive galaxy clusters} 

\author{
  E. S. Cypriano\altaffilmark{1,2} 
  L. Sodr\'e Jr. \altaffilmark{3}
  J.-P. Kneib \altaffilmark{4}
  and L. E. Campusano \altaffilmark{5}}

\altaffiltext{1}{SOAR Telescope, Casilla 603, La Serena, Chile
(esc@star.ucl.ac.uk).}

\altaffiltext{2}{Laborat\'orio Nacional de Astrof\'{\i}sica, CP 21, 37500-000,
Itajub\'a, MG, Brazil.}

\altaffiltext{3}{Departamento de Astronomia, Instituto de Astronomia,
Geof\'{\i}sica e Ciências Atmosf\'ericas, Universidade de S\~ao Paulo,
Rua do Mat\~ao 1226, Cidade Universit\'aria, 05508-090 S\~ao Paulo, Brazil
(laerte@astro.iag.usp.br).}

\altaffiltext{4}{Laboratoire d'Astrophysique de Marseille, Traverse du Siphon,
B.P.8 13376 Marseille Cedex 12, France (jean-paul.kneib@oamp.fr).}

\altaffiltext{5}{Universidad de Chile, Departamento de Astronom\'{\i}a, Casilla
  36-D, Santiago, Chile (luis@das.uchile.cl).}

\shortauthor{Cypriano et al.}
\shorttitle{Dynamical state of clusters}

\listofauthors{E. S. Cypriano, L. Sodr\'e Jr., J.-P. Kneib  \& L. E. Campusano}
\indexauthor{Cypriano, E. S.}
\indexauthor{Sodr\'e Jr., L.}
\indexauthor{Kneib, J.-P.}
\indexauthor{Campusano, L. E.}

\abstract{
We study the mass distribution of a sample of  24 X-ray bright
Abell clusters through weak gravitational lensing. This method is independent
of the  dynamical state of the galaxy cluster. Hence, by comparing dynamical
and lensing mass estimators, we can access the dynamical state of these
clusters. We have found that clusters with ICM temperatures above 8 keV show
strong deviations from the relaxation, as well as the presence of prominent
sub-structures. For the remaining clusters (the majority of the sample) we have
found agreement among the several mass estimators, which indicates that most of
the clusters are in or close to a state of dynamical equilibrium.  
}

\resumen{Se eval\'ua la distribuci\'on de masa para 24 c\'umulos de Abell
brillantes en rayos-X usando el efecto debil de los lentes gravitacionales.
Este m\'etodo es independiente del estado din\'amico del c\'umulo, y por lo
tanto la comparacion con masas ``dinamicas'' posibilita una determinaci\'on del
estado din\'amico de los c\'umulos. Los c\'umulos con temperaturas ICM mayores
que 8 keV presentan desviaciones con respecto a los cumulos dinamicamente
relajados, como tambi\'en presentan sub-estructuras significativas. Los distintos
indicadores de masa dan valores consistentes entre si para los dem\'as (la
mayor\'{\i}a de los c\'umulos de la muestra), lo cual es indicativo de estados
cercanos al equilibrio din\'amico.}

\addkeyword{Galaxy Clusters}
\addkeyword{Gravitational Lensing}

\begin{document}
\maketitle

\section{Introduction}

There are several ways to measure the masses of galaxy clusters but the most
widely used methods are the dynamical ones, for instance the study of the 
line-of-sight velocity distribution of the member galaxies and the X-ray
measurements of the temperature and distribution of the intra-cluster gas. They
relay on the assumption that galaxies and/or the intra-cluster gas are reliable
tracers of the potential well or that all cluster components are in a state
dynamical equilibrium. Gravitational lensing methods, on the other hand, are
completely independent of the cluster dynamical state.

The goal of this study is to take advantage of both kind of methods to
identify whether a cluster is in dynamical equilibrium or not by comparing
indicators of the dynamical mass: the velocity dispersion
of the galaxies ($\sigma_v$) and the X-ray measured gas temperature
(T$_X$) with  weak-lensing  equivalent indicators.
Disagreement between dynamical and non-dynamical mass indicators can be
interpreted as an indication that the assumption of dynamical equilibrium is
not valid.
 
\section{Results}  
In \citet{eu} we have fitted the weak-shear data with singular isothermal
profiles (spherical and elliptical) from a sample of 24 Abell clusters with
X-ray luminosities higher than $5\times 10^{44}$ h$_{50}^{-2}$ erg s$^{-1}$. 
Here we will
present the results of the sample members with independent measurements of
either T$_X$ or $\sigma_v$.
The mass related parameter that came out
from these fits is an equivalent of the line of sight velocity dispersion
($\sigma_{SIS}$ or $\sigma_{SIE}$, respectively for spherical and elliptical
profiles). Assuming energy equipartition between cluster 
galaxies and gas, we obtain  $T_{SIS}$ (or $T_{SIE}$) through the relation:
$\sigma^2 = {k T / \mu m_H}$, where $\mu=0.61$ is the mean molecular weight
of the gas, $m_{H}$ is the hydrogen mass, and $k$ is the Boltzmann constant. 
In Figure 1 we compare the actually measured dynamical
mass indicators with their weak-lensing counterparts.

\begin{figure*}
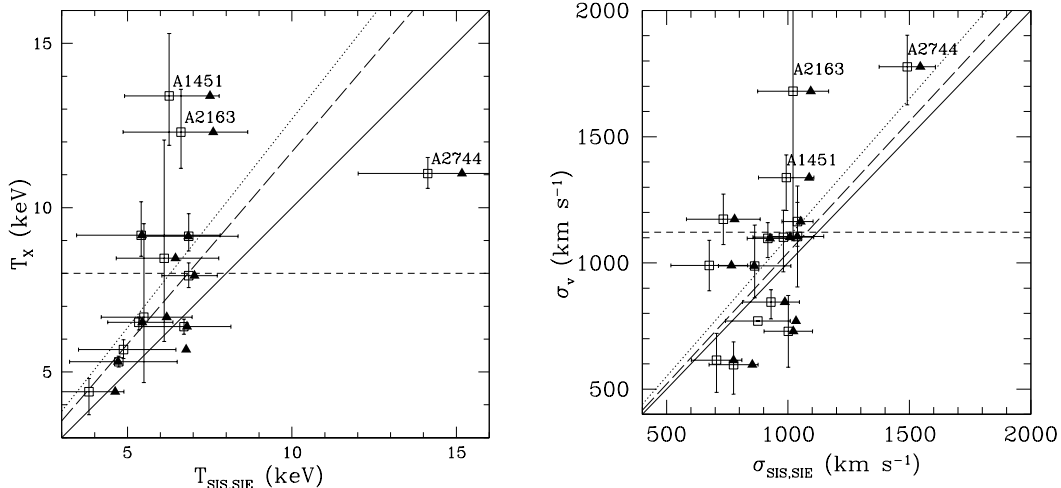

\begin{center}
\begin{tabular}{lr}
\includegraphics[width=0.85\columnwidth]{cypriano_e_fig1a.eps} &
\includegraphics[width=0.85\columnwidth]{cypriano_e_fig1b.eps}\\
\end{tabular}
\end{center}
\vspace{-0.8cm}
\caption{Comparison between the actually measured ICM temperatures (left panel) and velocity 
dispersions (right panel) with those inferred  by the fitting of
isothermal profiles (spheric--SIS and elliptical--SIE) to the shear data.
The squares correspond to the spherical model and
triangles to the elliptical. The error bars of the latter were suppressed 
for clarity. The solid line is defined by T$_{SIS,SIE}$ = \tx or
$\sigma_{SIS,SIE}$ = $\sigma_v$.
The dotted and long dashed lines show the best-fit  obtained 
with the SIS and SIE models, respectively, when the origin is kept constant.
The short dashed line indicates T$_X$ = 8 keV. 
Clusters with higher temperatures show signals of dynamic activity. 
\label{temp}}
\end{figure*}

It can be noted in Figure 1 that for most clusters 
the weak-lensing results agree with the dynamical data 
within 1.5 $\sigma$. However, this is not true for the entire sample.
For clusters with $T_X < 8$ keV (or equivalently $\sigma_v <1122$ \kms) the ratio
between the dynamical estimator and its weak-lensing equivalent is consistent
with the expected value of 1. Nonetheless, for clusters with $T_X > 8$ keV
strong disagreements can be found.
Among these cases three clusters stands out: A2744, A1451 and
A2163 (all labeled in the plots).  These clusters show temperatures and
velocity dispersions significantly different from the lensing estimations,
suggesting that they should be dynamically active.  Actually, detailed
individual analysis of these clusters provide support for this conclusion.

The study of the A2163 temperature map made by
\citet{Markevitch-Vikhlinin} with Chandra data shows at least two shocked
regions and other evidences that the central region of this cluster is in a
state of  violent motion.  In the same way, \citet{valtchanov}
describes A1451 as being in the final stage of establishing equilibrium after a
merger event, whereas its high X-ray temperature (13.4 keV) would be
probably due to a shock occurred recently.
A2744 seems to be an exception, since, contrarily to  A1451 and A2163,
which have both dynamical indicators in excess compared to the lensing
estimates, this cluster has a temperature  significantly lower
than that inferred from weak-lensing, but a higher  velocity dispersion.

The case of A2744 can be understood if, as suggested by the dynamical analysis
of \citet{girardi}, there are two structures along the line-of-sight.  
In this case $\sigma_v$ is artificially increased upwards, since it has
been measured over two superimposed galaxy velocity distributions.
Gravitational lensing is sensitive to the mass projected on the plane of the
sky, therefore its result is a weighted sum of the mass of both structures. 
Finally, T$_X$, which is obtained through X-ray spectroscopy, will be biased
in favour  of just one of the components, the brightest in X-rays,
giving a mass estimation lower than that obtained with the other two methods.
  
 In the hierarchical scenario, where most massive structures are formed through
the merger and accretion of less massive ones, massive galaxy clusters 
should be
forming at the present epoch. The results presented here reinforces this idea
by detecting dynamical activity among such systems, but not in the
less massive ones. One of the many consequences of this result is 
that massive clusters can  be
used as cosmological tools only if extra care is taken. Otherwise the 
presence of non-relaxed structures can bias the results.


\begin{thebibliography}

\bibitem[Cypriano et al.(2004)]{eu} Cypriano, E. S., Sodr\'e Jr., L.,
Kneib, J.-P. \& Campusano, L. E. 2004, ApJ, 613, 95

\bibitem[Girardi \& Mezzeti(2001)]{girardi} Girardi, M  \& Mezzeti, M. 2001, 
\apj, 548, 79

\bibitem[Markevitch \& Vikhlinin(2001)]{Markevitch-Vikhlinin} Markevitch, M.
\& Vikhlinin, A. 2001, \apj, 563, 95

\bibitem[Valtchanov et al.(2002)]{valtchanov} Valtchanov, I., Murphy, T.,
Pierre, M., Hunstead, R. \& L\'emonon, L. 2002, \aap, 392, 795

\end{thebibliography}
\end{document}